\documentclass[conference]{IEEEtran}
\IEEEoverridecommandlockouts

\usepackage{cite}
\usepackage{amsmath,amssymb,amsfonts}
\usepackage[ruled]{algorithm}
\usepackage[noEnd]{algpseudocodex}
\usepackage{graphicx}
\usepackage{textcomp}
\usepackage{pdfpages}
\usepackage{multirow}
\usepackage{subcaption}
\usepackage{xurl}
\PassOptionsToPackage{hyphens}{url}\usepackage{hyperref}
\usepackage{xcolor}
\def\BibTeX{{\rm B\kern-.05em{\sc i\kern-.025em b}\kern-.08em
    T\kern-.1667em\lower.7ex\hbox{E}\kern-.125emX}}

\usepackage{titlesec}

\titleformat{\subsubsection}[runin]
  {\normalfont\normalsize\itshape}
  {\thesubsubsection}{1em}{}[]

\usepackage{listings}

\lstdefinestyle{cuda}{
  language=C++, 
  basicstyle=\ttfamily\footnotesize, 
  keywordstyle=\color{blue}\bfseries, 
  commentstyle=\color{green!50!black}\itshape, 
  stringstyle=\color{orange}, 
  numbers=left, 
  numberstyle=\tiny\color{gray}, 
  stepnumber=1, 
  numbersep=5pt, 
  backgroundcolor=\color{gray!10}, 
  frame=single, 
  breaklines=true, 
  tabsize=2, 
  morekeywords={__global__, __device__, __host__, __shared__, __syncthreads, fmaxf, __shfl_up}, 
  morecomment=[l][\color{magenta}]{//}, 
  morecomment=[s][\color{magenta}]{/*}{*/}, 
  morestring=[b]" 
}

\begin{document}

\title{TrioSeq: A Novel Approach to Accelerate Triplet Sequence Alignment on GPUs}

\author{\IEEEauthorblockN{Miguel Graça}
\IEEEauthorblockA{
\textit{INESC-ID, Instituto Superior Técnico}\\
Lisbon, Portugal \\
miguel.graca@inesc-id.pt}
\and
\IEEEauthorblockN{Aleksandar Ilic}
\IEEEauthorblockA{
\textit{INESC-ID, Instituto Superior Técnico}\\
Lisbon, Portugal \\
aleksandar.ilic@inesc-id.pt}
}

\maketitle

\begin{abstract}
State-of-the-art multiple sequence alignment (MSA) algorithms are based on progressive approaches that rely on pairwise sequence alignment (PSA) to generate guide trees to align all sequences. Given an evidenced explosion in genomic data availability, research efforts have focused on accelerating PSA on massively-parallel architectures (e.g., GPUs) and specialized hardware (e.g., FPGAs). However, there is increasing evidence that starting from exact 3-way alignments could provide more robust, accurate MSAs, and improve genomic analysis. While the current literature has shown that PSA algorithms can be extended to align sequence triplets, the existent state-of-the-art on hardware acceleration of exact 3-way alignments is still scarce. In particular, current GPU methods are still inefficient due to lacking support for novel hardware features (e.g., cross-thread intrinsics), while being closed-source and vendor-specific. In this paper, TrioSeq is proposed as a fine-grained strategy to efficiently implement 3-way alignments on GPUs, leveraging novel levels of GPU parallelism and synchronization to achieve high throughput in aligning sequence triplets. Evaluation on NVIDIA and AMD GPUs shows that TrioSeq outperforms state-of-the-art GPU progressive methods on 3-way alignment by at least 20\% on simulated genomic datasets.

\end{abstract}


\section{Introduction}
\label{sec:Introduction}
The development of Next-Generation Sequencing (NGS) technologies \cite{hu2021next} represented a scientific breakthrough in DNA and protein sequencing, leading to an explosion of genomic data and widespread availability through open source projects (e.g., the 1000 Genomes Project \cite{10002015global} and the Genome In A Bottle (GIAB) consortium \cite{xiao2014giab}). For genomic analysis, one of the core algorithms in bioinformatics pipelines is sequence alignment, a fundamental technique used in genome assembly \cite{edgar2022petabase}, phylogenetic analysis \cite{ogden2006multiple}, and protein structure prediction \cite{yan2013comparative}. Sequence alignment aims to find an optimal order-preserving mapping of characters in $s$ sequences (composed of DNA or protein information), while allowing for possible character insertions, deletions, or gaps.

The case $s = 2$ is known as pairwise sequence alignment (PSA), which has been studied significantly in the literature \cite{prousalissurvey}. The most popular algorithms for PSA are Needleman-Wunsch \cite{needleman1970general} (for global alignments) and Smith-Waterman \cite{smith1981identification} (for local alignments) algorithms, which rely on dynamic programming (DP) to achieve optimal alignment scores, with $\mathcal{O}(n^2)$ complexity. When aligning $s$ sequences, with $s \geq 3$, a problem commonly known as multiple sequence alignment (MSA), both time and memory requirements for an optimal DP algorithm scale with $\mathcal{O}(n^s)$, although it is possible to reduce memory to $\mathcal{O}(n^{s-1})$ \cite{just2001computational, carrillo1988multiple}. Therefore, it is not feasible to align more than a few sequences of reasonable length due to the exponential complexity of exact MSA. 

As a consequence, MSA algorithms are based on progressive alignment algorithms \cite{maiolo2018progressive}, where PSA is a key operation to calculate all possible pairwise alignments between $s$ sequences (i.e., $\binom{s}{2}$ alignments are performed). The alignment scores are used as a distance matrix to create a phylogenetic guide tree, where closer sequences (i.e., with better alignment score) are clustered together \cite{thompson1994clustal}, using, for example, the UPGMA \cite{kaur2024construction} or neighbor-joining \cite{irawan2015construction} algorithms. The guide tree decides the order of sequences to align, starting by the closest two sequences. Therefore, progressive MSAs reduce the algorithm's complexity to $\mathcal{O}(n^2)$ (which is the complexity of PSA) at the cost of accuracy, as the optimal alignment is never calculated \cite{notredame2007recent}.

However, progressive MSA algorithms suffer from major drawbacks. First, when sequences are added to the guide tree, information from other sequences is not considered \cite{thompson1994clustal, wallace2006m}. Second, initial errors and gap placements are propagated to the final result \cite{seo2022correlations,loytynoja2008phylogeny}. To mitigate these problems, aligning sequence triplets ($s = 3$) has been shown to be advantageous in practice \cite{kruspe2007progressive} to improve gap placement on MSAs \cite{rosenberg2005multiple} and generate better guide trees \cite{colbourn2007lower}, which lead to more robust MSAs. Additionally, exact 3-way alignment improves the accuracy of MSAs when sequences are highly divergent \cite{askari2024three}, which has an impact on phylogeny reconstruction \cite{capella2013measuring,noah2020major} and epidemiological analysis \cite{lambert2019impact, llados2021accurate}. Furthermore, it has been shown that the Needleman-Wunsch algorithm for PSA can be extended to $s = 3$ \cite{gotoh1986alignment} to achieve the optimal alignment of sequence triplets. However, the cubic complexity of the algorithm, as well as the size of real-world NGS datasets, which are typically composed of many short-reads (sequences that span a few hundred base-pairs, typically generated by Illumina technology \cite{heydari2018browniealigner}), warrants the need for a fast and efficient solution.

Current research efforts have focused on accelerating PSAs on CPUs \cite{daily2016parasail, misra2018performance, rahn2018generic}, GPUs \cite{de2016cudalign, muller2022anyseq, ahmed2019gasal2, awan2020adept, schmidt2024cudasw++}, IPUs \cite{burchard2023space, zhao2024ipuma}, and FPGAs \cite{fei2018fpgasw, haghi2021fpga, rucci2018swifold,ben2020block} to boost the performance of progressive MSAs. Works on accelerating exact triplet alignment are more scarce and limited to a few approaches based on CPUs \cite{bani2024accelerating}, GPUs \cite{li2014optimal, sundfeld2013msa} and, more recently, FPGAs \cite{chien2018three, chien2022traceback}. In particular, GPU approaches for the exact alignment of sequence triplets suffer from three major drawbacks. First, while the state-of-the-art GPU approaches may have explored the maximum capabilities of GPUs for exact 3-way alignment, the landscape on the most recent hardware features (e.g., cross-thread intrinsics, shared memory) remains unexplored and may prove essential to fully exploit the hardware and achieve unprecedented performance in aligning sequence triplets with exact algorithms. Second, the existing solutions only target the global alignment of triplets, although semi-global and local alignments are also fundamental to detect shared motifs \cite{frith2004finding} in the sequences. Third, performing a progressive alignment of $s$ sequences based on aligning triplets would be equivalent to processing $\binom{s}{3}$ alignments, which, for large $s$, leads to a significant time-to-solution for a single device. Therefore, an efficient, distributed approach that leverages supercomputing platforms is fundamental to analyze large-scale triplet alignments in a reasonable time.

This is a particular gap that this work intends to close by proposing TrioSeq\footnote{Source code and datasets available at: https://github.com/champ-hub/trioseq-align}, an efficient GPU implementation to align three sequences using the extended Needleman-Wunsch algorithm that leverages cross-thread intrinsics, shared memory, and distributed shared memory to accelerate alignment computations while minimizing global memory access. The main contributions of this work are as follows:

\begin{enumerate}
    \item An efficient, fine-grained massively parallel exact 3-way local, semi-global, and global alignment approach to achieve high performance on GPUs.
    \item Scalability in single- and multi-GPU environments and portability across different devices.
\end{enumerate} 

Contribution (1) is concerned with the efficient mapping of the 3-way alignment algorithm to exploit massive data parallelism. To achieve high performance, the proposed solution focuses on the use of cross-thread intrinsics and shared memory to efficiently map the alignment computations to the hardware, which are implemented in CUDA and HIP to target GPUs from major vendors (NVIDIA and AMD). In the particular case of CUDA, thread block clusters (a recent hardware feature from H100) are also leveraged to further boost the proposed approach's scalability. Furthermore, to the best of our knowledge, it is the first time that exact semi-global and local 3-way alignment is implemented on GPUs.

Contribution (2) focuses on testing the proposed approach's portability and scalability across different devices. Compared with state-of-the-art GPU-based approaches for multiple sequence alignment (namely TWILIGHT \cite{tseng2025ultrafast} and CUDA ClustalW \cite{hung2015cuda}), the results on simulated datasets show that TrioSeq outperforms all works by at least 20\% in execution time, with a 11x speedup on datasets generated from real genomic data. Furthermore, an extensive evaluation in simulated sequence triplets shows that TrioSeq's alignments exhibit reduced false positive and false negative rates in 98\% of the tests. Finally, tests on two different supercomputing platforms demonstrate that TrioSeq achieves near-perfect scalability up to 64 GPUs for various simulated dataset distributions. This is achieved by exploring two static work distributions of a dataset (blocked and interleaved), as well as one dynamic work distribution (based on the amount of computation that each sequence triplet requires) to study scalability in supercomputing platforms, which are essential to process real-world large-scale datasets, leveraging distributed systems for exact 3-way alignments for the first time.

This work is structured as follows. Section \ref{sec:Background} provides the necessary background to understand the alignment algorithms for sequence triplets. Section \ref{sec:Methods} focuses on the algorithmic details to map exact 3-way alignments to GPU hardware. In Section \ref{sec:Results}, the proposed approach is evaluated on benchmark datasets, compared with state-of-the-art solutions, and tested on multi-GPU environments to assess scalability. Section \ref{sec:Related} highlights the main state-of-the-art approaches for MSAs and exact triplet alignments. Section \ref{sec:Conclusions} discusses the key takeaways from the paper and provides future research lines for exact 3-way alignments.

\section{Background}
\label{sec:Background}

\begin{figure}[t]
\centering
\includegraphics[width=\linewidth]{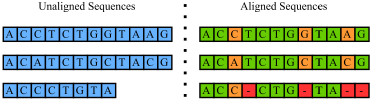}
\caption{Example of a 3-way alignment.}
\label{fig:alignment}
\end{figure}

Sequence alignment aims to identify the similarity between different sequences by matching the characters and adding gaps to describe possible character insertions or deletions in one or more sequences. To perform this calculation, sequence alignment datasets provide sets of unaligned $s$ sequences (typically DNA, RNA, or protein sequences) in text format. Figure \ref{fig:alignment} provides an example for $s = 3$, highlighting existent matches (green), mismatches (orange), and gaps (red) in alignment due to deletions in the last sequence.

To perform sequence alignment, Needleman-Wunsch \cite{needleman1970general} and Smith-Waterman \cite{smith1981identification} are standard DP algorithms, tailored for PSA ($s = 2$), which fill a two-dimensional grid and optimize the alignment score by adding gaps to both sequences to maximize character similarity. To align sequence triplets ($s = 3$), as demonstrated by Gotoh \cite{gotoh1986alignment}, it is possible to extend the algorithm's definition to a three-dimensional grid and, therefore, calculate the exact alignment of sequence triplets. Given three sequences, $S_0$, $S_1$, and $S_2$, of lengths $a$, $b$, and $c$, the 3-way alignment of the sequences, assuming a linear gap penalty (i.e., all gaps have a constant value), corresponds to the construction of a three-dimensional tensor using a DP algorithm defined by the recurrences given below

{\small
\begin{equation}
  M_{ijk} = max 
    \begin{cases}
      M_{i-1,j-1,k-1} + SOP(S_0[i-1], S_1[j-1], S_2[k-1])\\
      M_{i-1,j-1,k} + SOP(S_0[i-1], S_1[j-1], -)\\
      M_{i-1,j,k-1} + SOP(S_0[i-1], -, S_2[k-1])\\
      M_{i,j-1,k-1} + SOP(-, S_1[j-1], S_2[k-1])\\
      M_{i-1,j,k} + SOP(S_0[i-1], -, -)\\
      M_{i,j-1,k} + SOP(-, S_1[j-1], -)\\
      M_{i,j,k-1} + SOP(-, -, S_2[k-1])\\
      v\\
    \end{cases}  
    \label{eq:algorithm}
\end{equation}
}

where $-$ represents a gap in the alignment (due to insertions or deletions) and $SOP$ denotes the sum-of-pairs \cite{ye2004approximate} operation, given by

{\small
\begin{align*}
    SOP(S_0[i-1], S_1[j-1], S_2[k-1]) & = \sigma(S_0[i-1], S_1[j-1])\\
    & + \sigma(S_0[i-1], S_2[k-1])\\
    & + \sigma(S_1[j-1], S_2[k-1])\\
\end{align*}
}
where $\sigma$ represents a scoring scheme dependent on the evaluated characters matching (equal characters), mismatching (different characters), or aligning with a gap (one character and one gap). If both characters are gaps, the SOP returns 0. The variable $v$ and the initialization of the tensor determine the type of alignment (global, semi-global or local).

\subsubsection*{Global Alignments}. These alignments always start at $M_{000}$ and end at $M_{abc}$, where the optimal alignment score is stored. For global alignments, $v$ is set to $-\infty$ and the initialization, using SOP, is given by
\begin{align*}
    M_{000} & = 0,\\ 
    M_{i00} & = 2 \times i \times \alpha, M_{0j0} = 2 \times j \times \alpha, M_{00k} = 2 \times k \times \alpha,\\
    M_{ij0} & = max \begin{cases}
      M_{i-1,j-1,0} + SOP(S_0[i-1], S_1[j-1], -)\\
      M_{i-1,j,0} + SOP(S_0[i-1], -, -)\\
      M_{i,j,0} + SOP(-, S_1[j-1], -),\\
    \end{cases}\\
    M_{i0k} & = max \begin{cases}
      M_{i-1,0,k-1} + SOP(S_0[i-1], -, S_2[k-1])\\
      M_{i-1,0,k} + SOP(S_0[i-1], -, -)\\
      M_{i,0,k-1} + SOP(-, -, S_2[k-1]),\\
    \end{cases}\\
    M_{0jk} & = max \begin{cases}
      M_{0,j-1,k-1} + SOP(-, S_1[j-1], S_2[k-1])\\
      M_{0,j-1,k} + SOP(-, S_1[j-1], -)\\
      M_{0,j,k-1} + SOP(-, -, S_2[k-1]),\\
    \end{cases}
\end{align*}
where $\alpha$ is the gap penalty value. The output is an end-to-end sequence alignment that aligns every character.

\subsubsection*{Semi-Global Alignments}. In semi-global alignments, gaps at the beginning or at the end of the alignment are not penalized. While $v$ remains unchanged from the global alignment specifications, the initialization changes to
\begin{equation*}
    M_{i00} = M_{0j0} = M_{00k} = 0
\end{equation*}
and the optimal score is determined by the last slice of the tensor in any direction (i.e., $M_{ajk}$, $M_{ibk}$, $M_{ijc}$). Semi-global alignments are designed to find the optimal match of one sequence within another (e.g., when short sequences have to be aligned to long references \cite{carroll2019semiglobal}).

\subsubsection*{Local Alignments}. These alignments can start and end at any position of the tensor. Therefore, the tensor initialization is the same as semi-global alignments, with $v = 0$ as an additional constraint (i.e., all values in the tensor will be non-negative). Local alignments find the best matching subsequences, which is ideal to find shared motifs across various sequences.

\section{Related Work}
\label{sec:Related}

\begin{figure}[t]
\centering
\includegraphics[width=\linewidth]{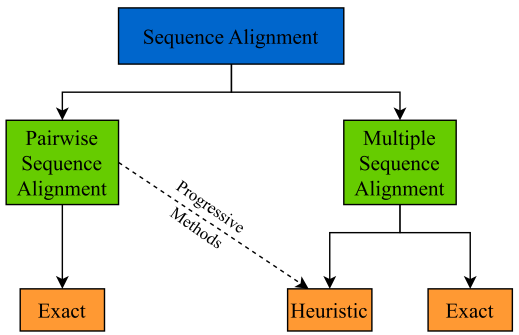}
\caption{State-of-the-art on sequence alignment.}
\label{fig:sota}
\end{figure}

The state-of-the-art on sequence alignment methods can be classified as shown in Figure \ref{fig:sota}. The existent approaches for PSA focus on the exact algorithms, while MSA tools are typically heuristic and mostly based on progressive methods that are built around PSA, with only a few approaches doing exact MSA, as will be discussed in this section.

\subsubsection*{Pairwise Sequence Alignment}. The parallelization of sequence alignment algorithms has focused primarily on the acceleration of PSAs to handle whole genome alignments and batches of different alignments of short sequences. Multicore processor approaches combine SIMD vectorization and multi-threading \cite{daily2016parasail, misra2018performance}, along with wavefront parallelism for intra-sequence parallelization \cite{hou2016aalign, liu2014swaphi}. To achieve a breakthrough in PSA performance, research efforts have focused on massively parallel architectures, such as GPUs \cite{nobile2017graphics, kong2025review} and, more recently, IPUs \cite{zhao2024ipuma, burchard2023space}.

Early GPU approaches compute PSAs at the level of a thread block and are optimized for specific use cases (e.g., protein database searches \cite{liu2013cudasw++}, genomic sequences \cite{de2016cudalign}). More recent solutions have focused on the development of flexible libraries, tailored for high performance, such as GASAL2 \cite{ahmed2019gasal2}, ADEPT \cite{awan2020adept}, NVBIO \cite{nvbio2015}, and AnySeq \cite{muller2020anyseq}. The current fastest PSA approaches are CUDA SW 4.0++ \cite{schmidt2024cudasw++} (for proteins) and AnySeq/GPU \cite{muller2022anyseq} (for DNA), which are the first solutions to process PSAs at warp level. However, these efficient frameworks exist as standalone tools for PSA, without integration in current MSA tools.

\subsubsection*{Progressive Multiple Sequence Alignment}. Among MSA progressive aligners, one of the most prominent works is the ClustalW framework \cite{thompson1994clustal}, which uses neighbor-joining and weighted SOP to derive the guide tree. To accelerate computations, further works have ported ClustalW to distributed systems using MPI \cite{li2003clustalw, de2021multi}, as well as GPUs \cite{oliver2005using} and FPGAs \cite{hung2015cuda}. Another framework, herein referred to as MAFFT \cite{katoh2010parallelization}, leverages multicore processors to accelerate all stages of MSA (i.e., all pairwise alignments, guide tree construction, and progressive alignment). More recently, TWILIGHT \cite{tseng2025ultrafast} emerged as a novel MSA tool that runs on CPUs and GPUs and is optimized for memory, speed, and accuracy for large MSAs by dividing the guide tree into subtrees, performing the alignments in parallel, and merging the subtree alignments.

\subsubsection*{Exact Triplet Sequence Alignment}. For the specific case of aligning triplets, the standard PSA algorithms can be extended to perform an exact 3-way sequence alignment \cite{gotoh1986alignment}, which have been shown to provide more accurate MSAs \cite{kruspe2007progressive, rosenberg2005multiple}. However, the extended algorithm scales in $\mathcal{O}(n^3)$ in time and memory, which warrants the need for fast, efficient approaches. MSA-GPU \cite{sundfeld2013msa} is one of the first approaches to implement optimal 3-way alignments on GPUs, which leverages the Carrillo-Lipman \cite{carrillo1988multiple} lower bound of MSAs to prune the search space and computes the 3D tensor in a multidimensional wavefront, although each thread calculates only one cell of the current wavefront. Another work \cite{li2014optimal} focuses on partitioning the three-dimensional tensor into chunks and assigning them to different SMs on a GPU, but the global memory transfers scale with the product of the sequences (i.e., $\mathcal{O}((|S_0| |S_1| |S_2|))$, which results in a I/O bottleneck. More recently, FPGA approaches have proposed efficient traceback implementations \cite{chien2022traceback, chien2018three} to retrieve the alignments from the 3D tensors. More recent works on exact MSA \cite{bani2024accelerating} tested three different parallel approaches for exact 3-way alignment, based on anti-diagonals and dividing the 3D tensor in blocks or slices, on multicore CPUs, but only on very short sequences.

While GPU solutions for 3-way alignment exist, these approaches do not leverage the current computational capabilities of GPUs and the most recent hardware features and libraries. The proposed algorithm, TrioSeq, is a novel approach to accelerate exact alignment of sequence triplets by leveraging hardware features of current GPU architectures, overcoming the limitations of the state-of-the-art techniques, as will be discussed in Section \ref{sec:Methods}).

\section{Proposed Methods}
\label{sec:Methods}
In this section, the proposed GPU-based approach to accelerate exact alignment of sequence triplets is described. We define TrioSeq to process sequence triplets at the level of a single GPU by using cross-thread intrinsics at the warp-level, shared memory for multiple warps, and distributed shared memory for multiple thread blocks to minimize global memory accesses and synchronization and maximize the amount of computation that individual threads perform to achieve high throughput for exact 3-way alignments.

\begin{figure}[t]
\centering
\includegraphics[width=\linewidth]{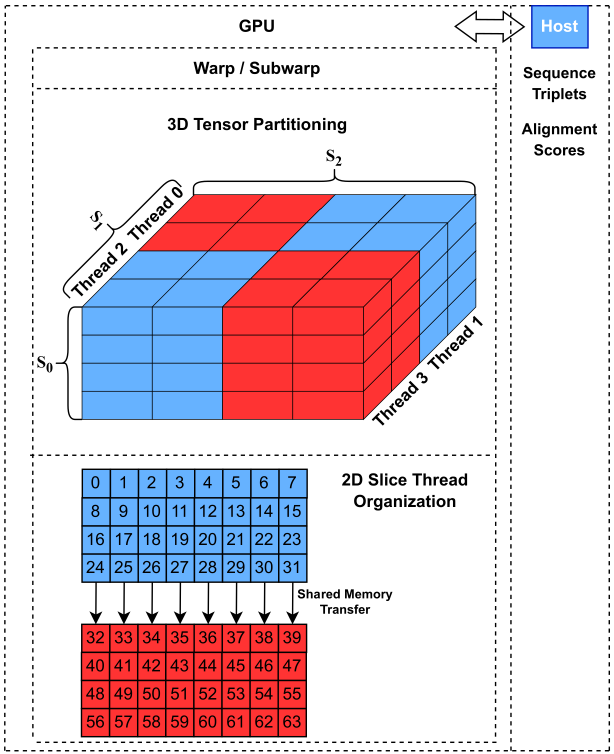}
\caption{Warp-Based Exact 3-Way Alignment.}
\label{fig:cube}
\end{figure}

\subsection{Approach Overview}

Figure \ref{fig:cube} provides a high-level overview of the TrioSeq approach. At the host level, sequence triplets are read from a dataset and sent to the GPU to perform the alignment computations, while the optimal scores are sent back to the host upon completion on the GPU. At the GPU level, TrioSeq calculates the 3D tensor that defines the alignment between the three sequences ($S_0$, $S_1$, and $S_2$) at the level of a warp of size $W$ (where $W = 32$ for NVIDIA architectures and $W = 64$ for AMD architectures). Processing alignments in warps has three advantages. First, cells from the 3D tensor can be calculated in thread registers, minimizing global memory load/store operations. Second, each thread can calculate multiple cells, which increases throughput. Third, cell values within a warp are exchanged by using low latency, cross-thread intrinsics, allowing for fast synchronization between threads. 

To fill the 3D tensor, each thread is assigned a chunk in row-major ordering (as an example, in Figure \ref{fig:cube}, 4 threads process a triplet alignment, with each thread responsible by one fourth of the tensor). To process each 3D chunk, all threads iterate over the first sequence's characters ($S_0$) to calculate a 2D $N \times N$ slice of its chunk in thread registers, while each individual thread reads $N$ characters from $S_1$ and $S_2$ (in Figure \ref{fig:cube}, $N = 2$). The number of threads that process an alignment, $T$, is a square number, and a complete 2D slice has a side of length $N \times \sqrt{T}$ (in Figure \ref{fig:cube}, $T = 4$ and $\sqrt{T} = 2$). After processing the full 3D tensor, the optimal alignment score is written to global memory and sent back to the host.

Due to the limitations of thread registers, the efficient calculation of an exact 3-way alignment at the level of a single warp becomes infeasible as sequence lengths increases, as $N$ would increase to keep a constant number of threads, which would result in register spilling and lower throughput. To maintain a high throughput, TrioSeq leverages multiple warps for bigger sequences and shared memory to exchange values between warps within the same thread block. In Figure \ref{fig:cube}, an example with $T = 64$ for $W = 32$ is provided. 
The last $\sqrt{T}$ threads (i.e., the last 8 threads) in the first warp store in shared memory the last row of their $(N+1) \times (N+1)$ square, which is accessed by the first $\sqrt{T}$ threads of the next warp, due to the row-major ordering of the threads to partition the complete 2D slice. For sequences that require a number of threads beyond the maximum allowed in a single thread block, thread block clusters are employed. Following the same strategy, the last $\sqrt{T}$ threads in a thread block store in distributed shared memory the last row of their $(N+1) \times (N+1)$ square, which is accessed by the first $\sqrt{T}$ threads of the next thread block.

\subsection{Cross-Thread Intrinsics}

\begin{figure}[t]
\centering
\includegraphics[width=\columnwidth]{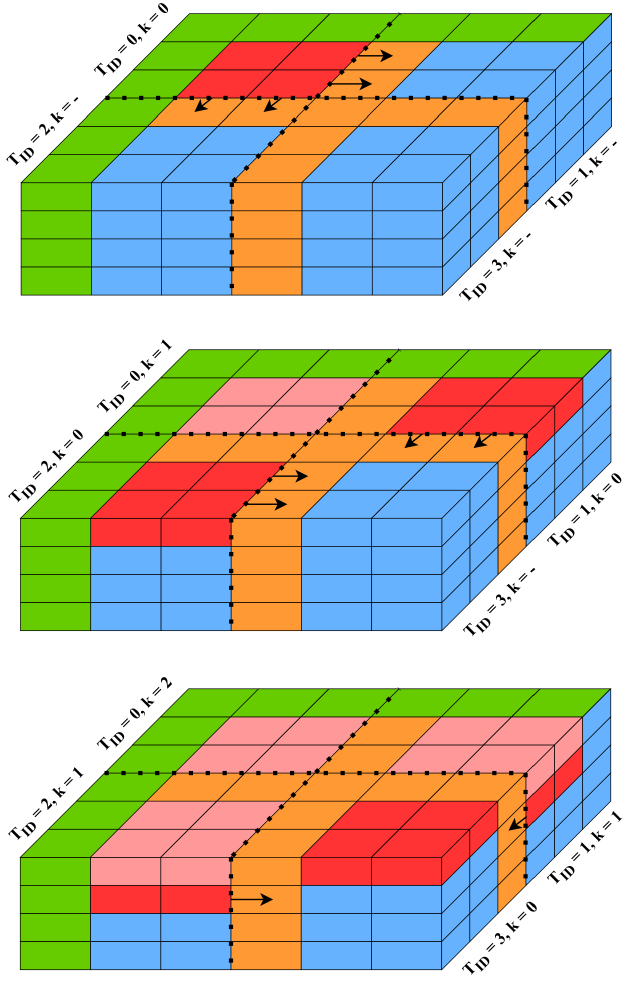}
\caption{General Pipeline for Triplet Sequence Alignment (for this example, $N$ = 2, with 4 threads per triplet).}
\label{fig:3way}
\end{figure}

TrioSeq's general approach, as described in the previous section, processes sequence triplets at the warp-level. To this end, a warp shuffle strategy is proposed to minimize global memory accesses and is illustrated in Figure \ref{fig:3way} with an example using 4 threads, with $N = 2$ for sequences of length 4. Each thread works on a $(N+1) \times (N+1)$ square and the algorithm performs an anti-diagonal sweep between threads, while each thread calculates its $N \times N$ square in a rowwise sweep.

In the first iteration (first cube), only thread 0 reads the first character of $S_0$ ($k = 0$) and calculates its $N \times N$ square (dark red cells) row by row, while the remaining cells (green) come from the initialization. After calculating the $N \times N$ cells (i.e., the red cells, using Equation \ref{eq:algorithm}), thread 0 sends its last column to thread 1 and its last row to thread 2 using warp shuffle instructions (which represent the orange cells in these threads). In the second iteration (second cube), thread 0 reads the second character of $S_0$ ($k = 1$), while threads 1 and 2 can now read the first character ($k = 0$) and calculate their individual $N \times N$ cells, using the values from the first column and first row, respectively, transferred from thread 0 (orange cells). Finally, in the third iteration (third cube), the last thread reads the first character ($k = 0$) and calculates its $N\times N$ cells.

Listing \ref{code:trioseq} describes the general algorithm for a warp-level global alignment, assuming $W = 32$ and using FP32. Line 2 provides the kernel arguments (i.e., the sequences, the scoring model, and the size of the subwarp/warp that is necessary to calculate the alignment). Line 4 defines the partial 2D slice that each thread calculates ($M$), line 5 defines auxiliary arrays, and lines 6-7 defines auxiliary variables. In line 8, the thread ID is used to locate the thread in the complete 2D slice. Line 9 defines arrays to store characters from $S1$ and $S2$. In line 12, the initialization stores the $N$ characters from $S1$ and $S2$ and initializes $M$, as well as the auxiliary arrays (line 5).

The main loop on $S0$ starts in line 14 and threads only perform computations when the condition on line 15 is true to ensure the anti-diagonal sweep is done. Lines 18-30 calculate the first row separately (to account for the threads that have initialization values on the first row of $M$) and store the new results on $R$. Lines 32-53 calculate the remaining cells of the $(N+1) \times (N+1)$ square, with lines 35-39 calculating elements of the first column separately (to account for the threads that have initialization values on the first column of $M$), and lines 41-50 calculate the remaining cells. In particular, lines 44-47 perform the alignment calculations according to Equation \ref{eq:algorithm}. Lines 55-61 store the values of $R$ (which has the new values for the last row of $M)$ in $M$. In lines 65-70, values are exchanged between threads by using shuffle intrinsics. Finally, in line 74, the result is written to global memory.

\begin{figure}[!t]
{\linespread{0.7}\selectfont
\begin{lstlisting} [ style=cuda, caption={GPU kernel for warp-level global alignment (FP32).}, label=code:trioseq]
template <unsigned int N, unsigned int P> 
__global__ void warp_global_fp32(char* S0, char* S1, char* S2, int subwarp, float Ma, float Mi, float gap, float* A)
{
  float M[N+1][N+1] = {{0}};
  float C[N+1] = {0}, R[N+1] = {0};
  float S_1 = S_2 = S_3 = 0;
  float y1 = y2 = y3 = y4 = 0;
  int IDC = threadIdx.x%subwarp;
  int IDR = (threadIdx.x/subwarp)%subwarp; 
  int idx = -IDC-IDR, offset = subwarp*2-1;
  char h1[N] = {0}, h2[N] = {0};
  char rs;

  //INITIALIZATION

  for(int r = 0; r < |S0| + offset; r++){
    if(idx >= 0 && idx < |S0|){
      rs = S0[idx];

      if(IDR == 0){
        if(IDC == 0) y1 = -2*(idx + 1)*gap;
        else y1 = C[0]; y4 = y1;

        for(int h = 1; h < N; h++){
          S_1 = (rs == h2[h-1]) ? Ma : Mi;
          y2 = fmaxf(y1, fmaxf(M[0][h], M[0][h-1] + S_1)) - 2*gap;
          R[h-1] = y1; y1 = y2;
        }

        if(IDC == 0) y4 = R[0];
        R[N-1] = y1;
      }

      for(int h = 1; h < N; h++){
        S_1 = (rs == h1[h-1]) ? Ma : Mi;
                
        if(IDC == 0){
          y1 = fmaxf(C[h], fmaxf(R[0], C[h-1] + S_1)) - 2*gap;
          C[h-1] = y4; y4 = y1;
        }
        else y1 = C[h]; y4 = y1;

        for(int l = 1; l < N; l++){
          S_2 = (rs == h2[l-1]) ? Ma : Mi;
          S_3 = (h1[h-1] == h2[l-1]) ? Ma : Mi;
          y3 = fmaxf(y1, fmaxf(M[h][l],R[l])) - 2*gap;

          y2 = fmaxf(M[h-1][l-1] + S_1 + S_2 + S_3, fmaxf(R[l-1] + S_3, fmaxf(M[h-1][l] + S_1, M[h][l-1] + S_2)) - 2*gap);

          M[h-1][l - 1] = R[l - 1]; 
          R[l-1] = y1; y1 = fmaxf(y2,y3);
        }

        M[h-1][N-1] = R[N-1]; R[N-1] = y1;
      }

      if(IDC == 0){
        C[N-1] = y4;
        M[N-1][0] = y4;
      }
      else M[N-1][0] = y4;

      for(int h = 1; h < N; h++) M[N-1][h] = R[h];
    }
    idx++;

    for(int h = 0; h < N; h++){
      y3 = __shfl_up(M[h][N-1], 1);
      if{IDC != 0} C[h] = y3;
      y3 = __shfl_up(M[N-1][h], subwarp);
      if{IDR != 0} R[h] = y3;
    }
        
  }
    
  if(IDR == sub-1 && IDC == sub-1) *A = M[N-1][N-1];
}
\end{lstlisting}
}
\end{figure}

To maximize performance, TrioSeq focuses on four different aspects. First, the value of $N$ must be carefully selected to maximize the number of cells that each thread can calculate individually while ensuring registers do not spill to local memory, which would increase execution time (the impact of $N$ is further discussed in Section \ref{sec:kernel}). Second, sequence triplets can be processed at multiple levels of parallelism. For small sequences, the 3-way alignment can be done at the level of a warp/subwarp. For example, if a single thread computes $N \times N$ elements of a 2D slice and the sequences have length $2N$, only 4 threads are necessary for the computations, which would allow 8 triplets to be processed if $W = 32$ or 16 triplets if $W = 64$. As the sequence length increases, multiple warps are leveraged, using shared memory to exchange values between warps, and, afterwards, using distributed shared memory to use multiple thread blocks and further boost scalability. Third, score computations are based on single precision arithmetic (FP32) and half precision arithmetic (FP16)\footnote{The maximum function to process two FP16 values (\textit{\_hmax2}) is implemented in CUDA, but not HIP, and is implemented manually for AMD GPUs.}, since the gap penalties and the value of $\sigma$ are typically below 10. By packing two FP16 as a FP32 value, two alignments are calculated simultaneously, which would improve alignment throughput by 2x. Finally, the proposed approach is scaled to multi-GPUs (as will be shown in Section \ref{sec:mpires}) to maximize the number of triplet alignments that can be processed at the same time.

\section{Experimental Results}
\label{sec:Results}
In this section, details on the experimental evaluation of the proposed sequence alignment framework are provided. First, we focus on the proposed approach's scalability and portability to answer the following questions: (i) how does the number of elements that each thread calculates (square of side $N$) impact the throughput?, (ii) how does the throughput scale with sequence length on different GPUs?, and (iii) how does the throughput scale in multi-GPU environments with different partitioning schemes? Second, the proposed approach is tested on simulated genomic datasets, with varying sequence lengths and sequence triplets to be processed, to study how TrioSeq compares with state-of-the-art GPU-based methods.

\subsection{Experimental Setup}

The reported experiments have been performed on 4 GPUs, i.e., NVIDIA A100 and H100, using the CUDA 12.8 toolkit, and AMD MI250X and MI300A, using the ROCM 6.0 toolkit. For distributed computing, the OpenMPI 5.0.3 library is used. Simulated alignment datasets of sequence triples with lengths between 8 and 1024 are generated using AliSim \cite{ly2022alisim} for benchmarking and comparing with state-of-the-art GPU approaches. The evaluation of TrioSeq and comparison to other state-of-the-art solutions focuses on the 3-way algorithm's throughput, measured as Tera Cell Updates Per Second (TCUPS) and calculated as

\begin{equation*}
    TCUPS = \frac{\sum_i^s (|S_0|_i \times |S_1|_i \times |S_2|_i)}{t \times 10^{12}},
\end{equation*}

where $t$ is the runtime in seconds, $s$ is the total number of sequence triplets to process, and $|S_0|_i$, $|S_1|_i$, and $|S_2|_i$ are the length of the $i$-th sequences.

\subsection{Kernel Analysis} \label {sec:kernel}

\begin{figure}[t] 
    \begin{subfigure}[t]{0.98\columnwidth}
    \includegraphics[width=\columnwidth]{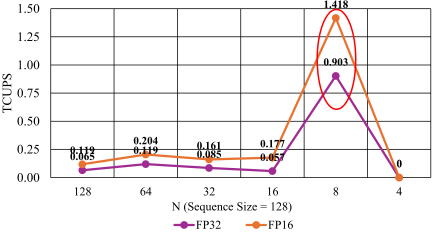}
    \caption{H100}
    \label{h100kernel}
    \end{subfigure}

    \begin{subfigure}[t]{0.98\columnwidth}
    \includegraphics[width=\columnwidth]{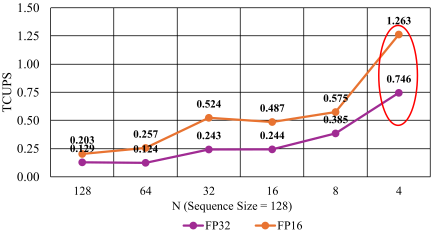}
    \caption{MI300A}
    \label{mi300kernel}
    \end{subfigure} 

    \caption{TCUPS evolution with the number of elements processed by each thread ($N$) in H100 (top) and MI300X (bottom), using FP16 and FP32.}
\end{figure}

In this section, an analysis on the impact of $N$ (the size of the $N \times N$ square that each thread calculates) in the proposed approach's performance is performed. 
It is worth mentioning that the size of $N$ impacts the amount of threads involved in the alignment execution. For example, for a sequence size of 128 and $N = 128$, a single thread calculates the complete alignment. Reducing $N$ to $64$ implies that the complete 2D slice is partitioned into 4 subslices of $64 \times 64$, which requires a total of four (4) threads to be involved in the computation. As such, for $N=16$ there are 64 threads involved, while for $N=4$ there are 1024 ($32 \times 32$) threads operating on the complete 2D alignment slice (the maximum amount of threads that can be run in a block in the considered GPU devices).

To this end, experiments on NVIDIA H100 and AMD MI300A are conducted for global alignment, fixing the sequence length to 128 and varying the value of $N$. Figures \ref{h100kernel} and \ref{mi300kernel} show the variations in TCUPS on H100 and MI300A GPUs, respectively, on FP32 and FP16, according to the value of $N$. It is worth noting that, in both devices, an equal amount of threads is involved in the alignment for a fixed value of $N$. However, the total warps differ among these two devices, i.e., it is doubled for NVIDIA H100 (since  NVIDIA H100 has a warp size of 32, while AMD MI300A has a warp size of 64).

If $N$ is small, the number of threads per block increases, and each thread performs little work, which decreases throughput. On the other hand, if $N$ is too large (in the limit, if a single thread calculates the complete tensor for a single triplet sequence alignment), the register pressure results in spilling to the global memory which, due to its high latency, hinders the kernel’s throughput. Therefore, one must achieve a balance between providing enough elements for threads to process while ensuring that register spilling does not occur. For example, Figure \ref{h100kernel} shows that the optimal value for H100 is $N = 8$. Note that the optimal value of $N$ is equal in FP16 and FP32, as registers are 32-bit and the two packed FP16 values are equivalent to a single FP32 value in memory size. The results also show that the optimal value of $N$ is hardware dependent and can be tuned to achieve maximum performance on different GPU architectures. For this work, A100 and H100 use $N = 8$, while MI250X and MI300A use $N = 4$, which correlates with the differences in the warp size. Furthermore, architectural differences in the way how register pressure is handled can also be observed in Figures \ref{h100kernel} and \ref{mi300kernel}, with AMD MI300A providing more gradual performance improvements. 

\subsection{Cross-vendor GPU Analysis}

\begin{figure*}[t] 
    \begin{subfigure}[t]{0.67\columnwidth}
    \includegraphics[width=\linewidth]{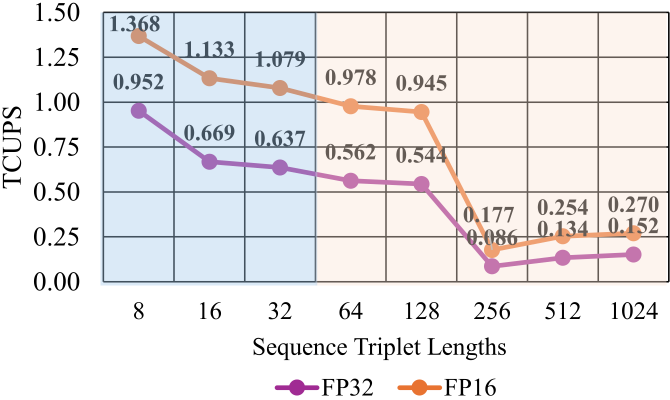}
    \caption{A100 (global)}
    \label{a100global}
    \end{subfigure}
    \begin{subfigure}[t]{0.67\columnwidth}
    \includegraphics[width=\linewidth]{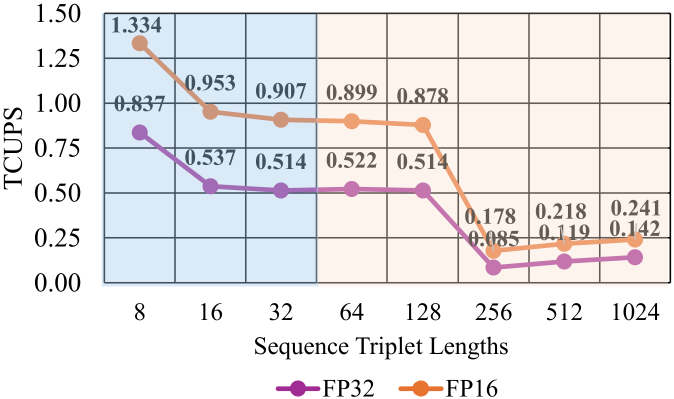}
    \caption{A100 (semiglobal)}
    \label{a100semiglobal}
    \end{subfigure} 
    \begin{subfigure}[t]{0.67\columnwidth}
    \includegraphics[width=\linewidth]{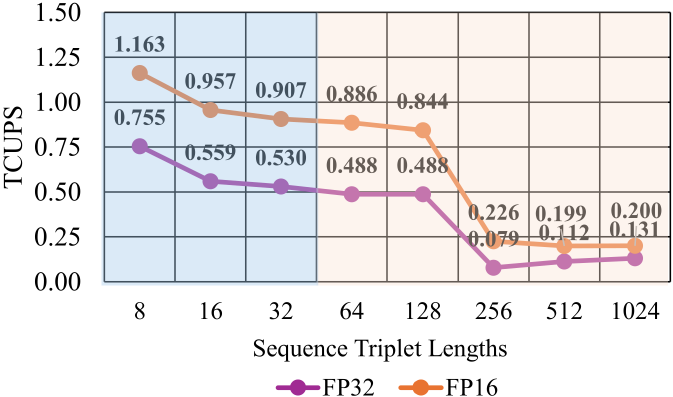}
    \caption{A100 (local)}
    \label{a100local}
    \end{subfigure} 

    \begin{subfigure}[t]{0.67\columnwidth}
    \includegraphics[width=\linewidth]{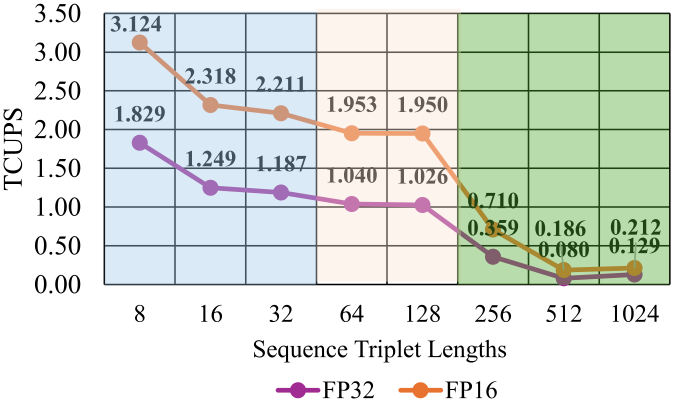}
    \caption{H100 (global)}
    \label{h100global}
    \end{subfigure}
    \begin{subfigure}[t]{0.67\columnwidth}
    \includegraphics[width=\linewidth]{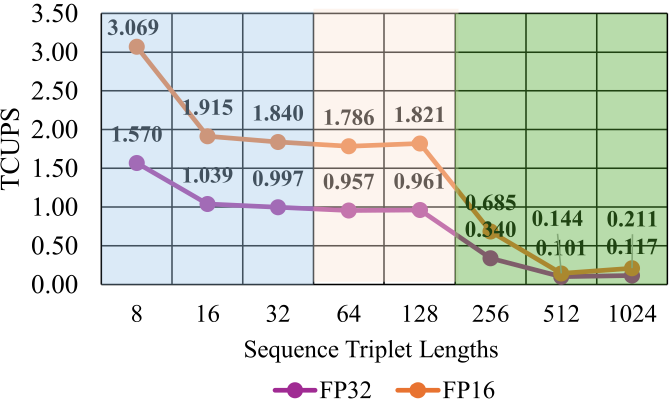}
    \caption{H100 (semiglobal)}
    \label{h100semiglobal}
    \end{subfigure} 
    \begin{subfigure}[t]{0.67\columnwidth}
    \includegraphics[width=\linewidth]{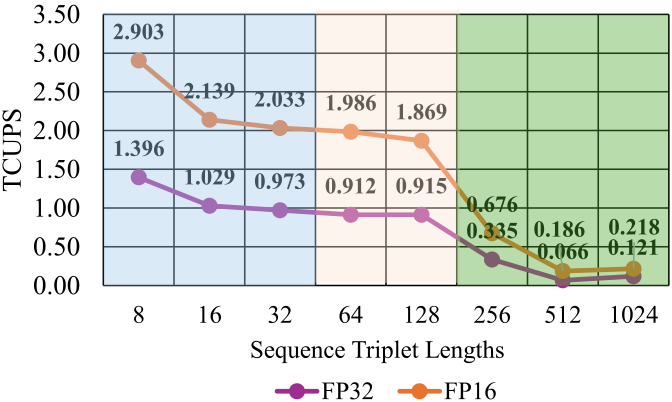}
    \caption{H100 (local)}
    \label{h100local}
    \end{subfigure}

    \begin{subfigure}[t]{0.67\columnwidth}
    \includegraphics[width=\linewidth]{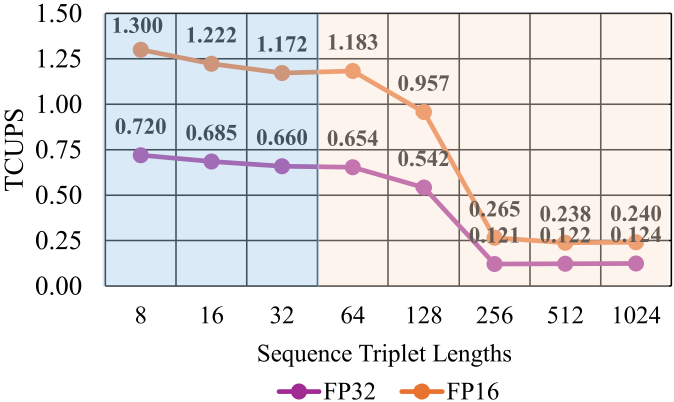}
    \caption{MI250X (global)}
    \label{mi250global}
    \end{subfigure}
    \begin{subfigure}[t]{0.67\columnwidth}
    \includegraphics[width=\linewidth]{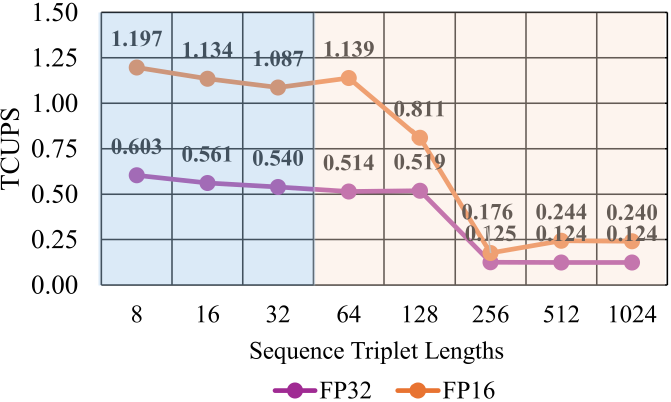}
    \caption{MI250X (semiglobal)}
    \label{mi250semiglobal}
    \end{subfigure} 
    \begin{subfigure}[t]{0.67\columnwidth}
    \includegraphics[width=\linewidth]{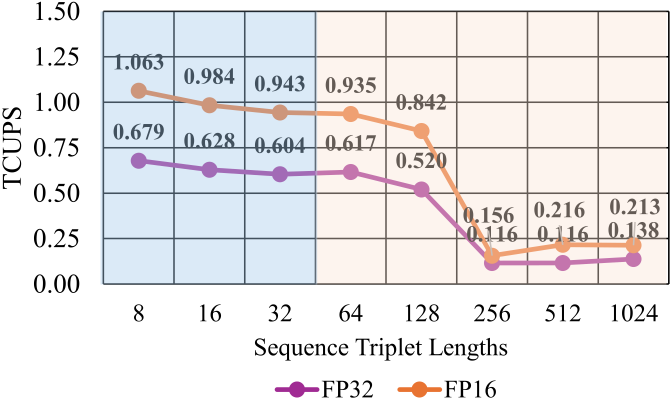}
    \caption{MI250X (local)}
    \label{mi250local}
    \end{subfigure} 

    \begin{subfigure}[t]{0.67\columnwidth}
    \includegraphics[width=\linewidth]{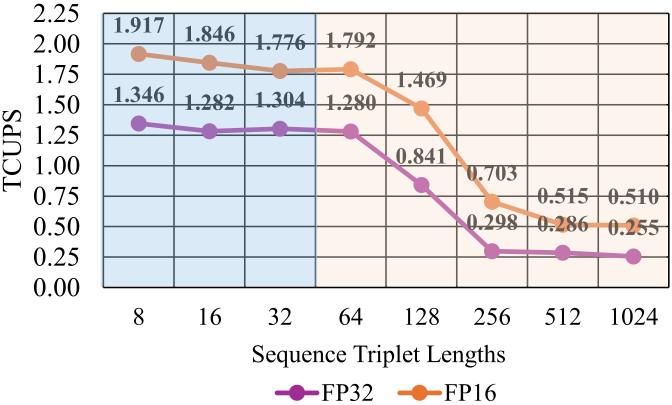}
    \caption{MI300A (global)}
    \label{mi300global}
    \end{subfigure}
    \begin{subfigure}[t]{0.67\columnwidth}
    \includegraphics[width=\linewidth]{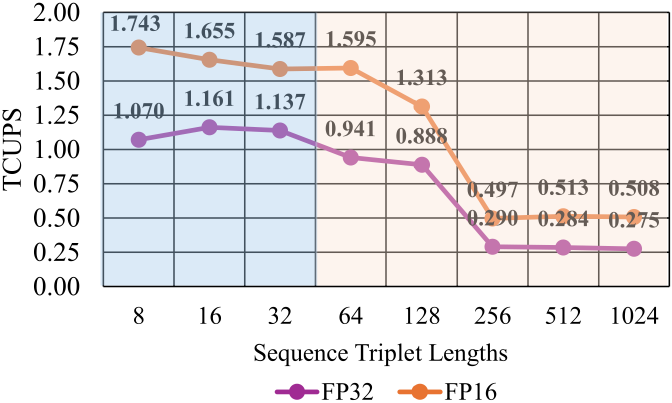}
    \caption{MI300A (semiglobal)}
    \label{mi300semiglobal}
    \end{subfigure} 
    \begin{subfigure}[t]{0.67\columnwidth}
    \includegraphics[width=\linewidth]{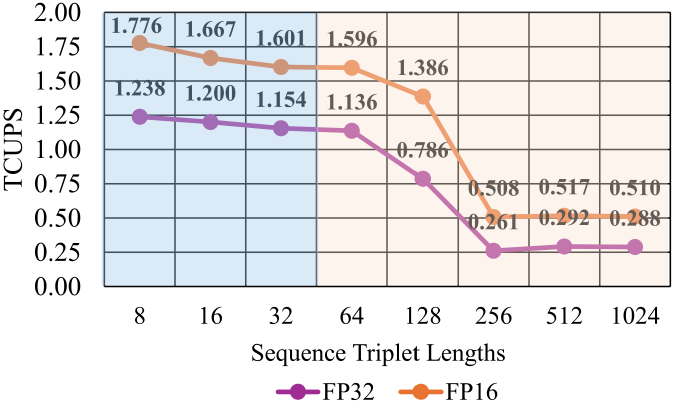}
    \caption{MI300A (local)}
    \label{mi300local}
    \end{subfigure} 

    \caption{TCUPS curves for different alignment algorithms in A100, H100, MI250X, and MI300A GPUs using FP16 and FP32.}
\end{figure*}

To evaluate TrioSeq's performance across different GPU architectures, the algorithms for exact triplet alignment are tested on the NVIDIA A100 and H100 GPUs, as well as the AMD MI250X and MI300A GPUs. Figures \ref{a100global} to \ref{mi300local} display the TCUPS evolution for FP16 and FP32 in A100, H100, MI250X, and MI300A, for the three possible alignments (global, semiglobal, and local).

In general, all tested GPUs follow the same trend, achieving high TCUPS values for small sequences, with throughput decreasing as the sequence lengths increases. The highest performance is achieved with a single warp execution (blue area), which slightly drops when multiple warps and shared memory are leveraged for the alignments (orange area), followed by the use of thread block clusters (green area). Among all tested devices, for all alignment types, H100 achieves the highest throughput for sequence lengths between 8 and 128 (between 1.8 and 3.1 TCUPS in FP16 and between 0.9 and 1.8 TCUPS for FP32). In all tested GPUs, for sequence lengths over 128, it is necessary to increase $N$ to avoid storing intermediate results in global memory. However, as mentioned in Section \ref{sec:kernel}, increasing $N$ results in register spilling to local memory, which reduces performance, explaining the lower throughput for sequence lengths beyond 128. Among the four evaluated GPUs, MI300A provides the best TCUPS for sequence lengths between 256 and 1024, achieving more than 0.5 TCUPS in FP16 and 0.25 TCUPS in FP32. Performance-wise, MI300A and H100 provide the best throughputs in TCUPS. In particular, H100 provides the best results for sequence lengths up to 256, while MI300A is better for longer lengths (512 and 1024). Therefore, the best suited device can be chosen depending on the range of sequence lengths of a given dataset, further reinforcing the need for efficient, cross-vendor solutions.

Overall, evaluating FP16, for global alignments, the highest TCUPS are 1.368 (A100), 3.124 (H100), 1.300 (MI250X), and 1.917 (MI300A), while for semi-global/local alignments, the peak TCUPS are 1.334/1.163 (A100), 3.069/2.903 (H100), 1.197/1.063 (MI250X), and 1.743/1.776 (MI300A). For FP32 and all three alignments, the peak TCUPS are 0.952/0.837/0.755 (A100), 1.829/1.570/1.396 (H100), 0,720/0.603/ 0.679 (MI250X), and 1.346/1.070/1.238 (MI300A). Note that the peak throughput for global alignment is typically higher, since semiglobal and local alignment algorithms require additional comparison operations. In particular, semiglobal alignment calculates the maximum value across the last slices of the alignment tensor, and local alignment requires comparing every value of the tensor with 0, which will decrease throughput.

Finally, to demonstrate the impact of the traceback on TrioSeq's peak throughput, a lightweight example of Hirschberg's algorithm with sequences of length 8 is tested. For FP32/FP16, the peak throughput for the traceback in global alignment is 0.541/0.875 TCUPS (A100), 0.921/1.660 TCUPS (H100), 0.343/0.625 TCUPS (MI250X), and 0.691/0.921 TCUPS (MI300A). For the scoring, as observed in Figures \ref{a100global}, \ref{h100global}, \ref{mi250global}, and \ref{mi300global}, the peak throughput for sequences of length 8 in FP32/FP16 is 0.952/1.368 TCUPS (A100), 1.829/3.124 TCUPS (H100), 0.720/1.300 TCUPS (MI250X), and 1.346/1.917 TCUPS (MI300A). Note that the traceback throughput is approximately half of the scoring throughput, which is in line with previous research on GPU implementations of Hirschberg's algorithm \cite{muller2022anyseq} for sequence alignment.

\subsection{Multi-GPU Evaluation} \label{sec:mpires}

\begin{figure}[t] 
    \begin{subfigure}[t]{0.99\columnwidth}
    \includegraphics[width=\columnwidth]{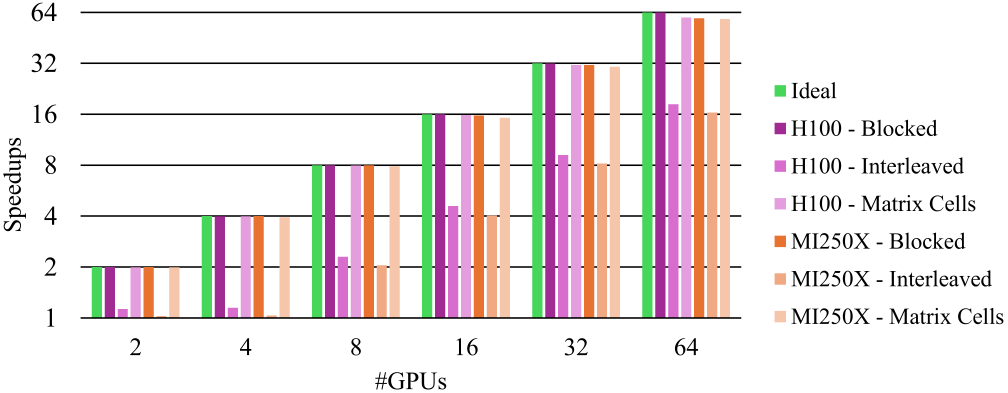}
    \caption{Blocked Dataset}
    \label{mpiblock}
    \end{subfigure}

    \begin{subfigure}[t]{0.99\columnwidth}
    \includegraphics[width=\columnwidth]{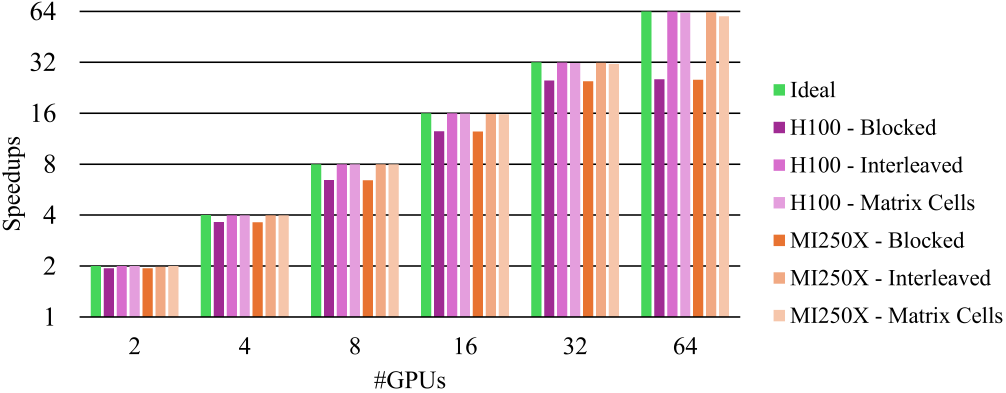}
    \caption{Interleaved Dataset}
    \label{mpiinter}
    \end{subfigure}

    \begin{subfigure}[t]{0.99\columnwidth}
    \includegraphics[width=\columnwidth]{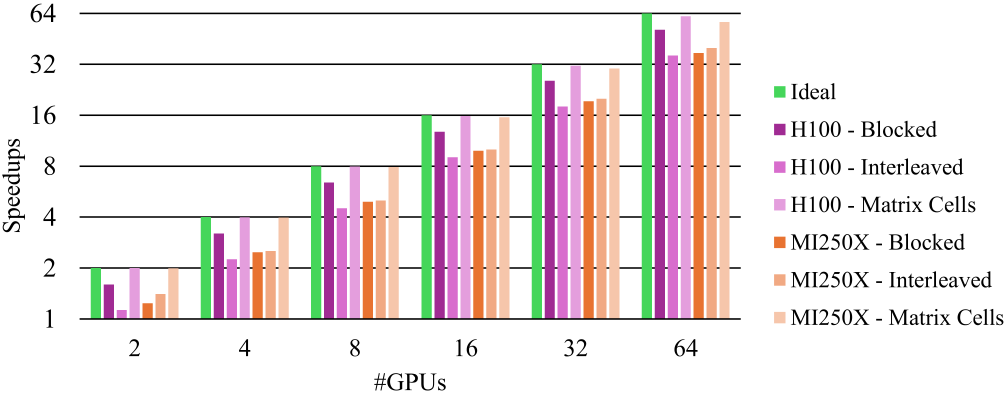}
    \caption{Random Dataset}
    \label{mpirandom}
    \end{subfigure}

    \caption{TrioSeq's scalability with the number of GPUs on different simulated datasets.}
\end{figure}

To assess TrioSeq’s scalability in distributed systems, experiments on H100 and MI250X are performed with different numbers of GPUs. To this end, two static and one dynamic partitioning approaches are studied. For the static partitioning, the first approach (blocked) divides the dataset in chunks and assigns each GPU one chunk for processing, while in the second approach (interleaved), sequence triplets are distributed in a round-robin fashion. For the dynamic partitioning, sequence lengths are taken into account for distributing the workload and each triplet is sent to the current GPU that is calculating the least cumulative matrix cells. Figures \ref{mpiblock} to \ref{mpirandom} show the scalability results with H100 and MI250X, respectively, up to 64 GPUs, for datasets with three possible distributions. The first dataset is ideal for blocked partitioning, the second is ideal for interleaved partitioning, and the third follows a random distribution of different sequences with different lengths.

For the first two datasets, the results show that near perfect scalability is achieved with the matching static partitioning scheme. Using 64 GPUs, the blocked partitioning scheme achieves a 63.9x/59.1x speedup on H100/MI250X on the first dataset, while the interleaved partitioning scheme on the second dataset achieves a 63.8x/63.2x speedup on H100/MI250X with the same number of GPUs. As expected, the use of interleaved partitioning on the blocked dataset or blocked partitioning in the interleaved dataset results in poor performance, with H100/MI250X achieving 18.4x/16.4x speedups on the former and   26.4x/25.3x speedups on the latter. In both cases, the proposed dynamic partitioning based on the amount of computation of each triplet achieves near perfect scalability with the number of GPUs. For the blocked/interleaved datasets, MI250X achieves a 58.5x/59.8x speedup with 64 GPUs, while H100 achieves a 59.7x/62.8x speedup.

For the third case, which processes a dataset with a random distribution of sequences, the results show the limitations of the evaluated static partitioning schemes, as H100/MI250X achieve 51.2x/37.3x speedups for the blocked distribution and 36.1x/40x speedups for the interleaved distribution, using 64 GPUs. However, the scalability of the work partitioning based on the sequence lengths to process is still near perfect, as H100/MI250X achieve 61.6x/56.9x speedups for 64 GPUs.

\begin{table}[t]
\setlength{\tabcolsep}{2pt}
\renewcommand{\arraystretch}{1}
\centering
{\fontsize{8pt}{10pt}\selectfont
\caption{Execution Times for TWILIGHT, CUDA ClustalW, and TrioSeq (in milliseconds).}
\label{tab:sota}
\begin{tabular}{c|lllll}
\hline
Length & GPU & TWILIGHT & CUDA ClustalW & TrioSeq & \textbf{Speedup}\\
\hline
\multirow{4}{*}{8} & A100 & 0.759 & 282.8 & 0.022 & \textbf{34.5x}\\
& H100 & 1.090 & 111.5 & 0.013 & \textbf{83.8x}\\
& MI250X & 0.708 & - & 0.045 & \textbf{15.7x}\\
& MI300A & 0.835 & - & 0.021 & \textbf{39,8x}\\
\hline
\multirow{4}{*}{16} & A100 & 0.744 & 296.8 & 0.042 & \textbf{17.7x}\\
& H100 & 1.075 & 115.9 & 0.028 & \textbf{38.4x}\\
& MI250X & 0.808 & - & 0.061 & \textbf{13.2x}\\
& MI300A & 0.980 & - & 0.036 & \textbf{27.2x}\\
\hline
\multirow{4}{*}{32} & A100 & 1.013 & 301.1 & 0.070 & \textbf{14.5x}\\
& H100 & 1.160 & 138.5 & 0.053 & \textbf{21.9x}\\
& MI250X & 0.836 & - & 0.101 & \textbf{8.28x}\\
& MI300A & 1.044 & - & 0.065 & \textbf{16.1x}\\
\hline
\multirow{4}{*}{64} & A100 & 1.646 & 303.1 & 0.160 & \textbf{10.3x}\\
& H100 & 1.628 & 140.8 & 0.100 & \textbf{16.3x}\\
& MI250X & 1.140 & - & 0.227 & \textbf{5.02x}\\
& MI300A & 1.429 & - & 0.137 & \textbf{10.4x}\\
\hline
\multirow{4}{*}{128} & A100 & 3.186 & 309.3 & 0.646 & \textbf{4.93x}\\
& H100 & 2,335 & 143.2 & 0.311 & \textbf{7.51x}\\
& MI250X & 2.547 & - & 0.954 & \textbf{2.67x}\\
& MI300A & 2.363 & - & 0.548 & \textbf{4.31x}\\
\hline
\multirow{4}{*}{256} & A100 & 6.831 & 316.3 & 4.331 & \textbf{1.58x}\\
& H100 & 5.786 & 144.6 & 1.432 & \textbf{4.04x}\\
& MI250X & 7.361 & - & 5.222 & \textbf{1.41x}\\
& MI300A & 7.010 & - & 3.618 & \textbf{1.94x}\\
\hline
\multirow{4}{*}{512} & A100 & 34.29 & 325.1 & 22.32 & \textbf{1.54x}\\
& H100 & 19.02 & 149.1 & 15.66 & \textbf{1.21x}\\
& MI250X & 26.16 & - & 20.89 & \textbf{1.25x}\\
& MI300A & 23.57 & - & 14.47 & \textbf{1.63x}\\
\hline
\multirow{4}{*}{1024} & A100 & 108.4 & 347.0 & 89.29 & \textbf{1.21x}\\
& H100 & 64.67 & 153.8 & 52.79 & \textbf{1.22x}\\
& MI250X & 106.3 & - & 83.55 & \textbf{1.27x}\\
& MI300A & 73.33 & - & 57.88 & \textbf{1.27x}\\
\hline
\end{tabular}
}
\end{table}

\subsection{State-of-the-art Comparison}

In this section, TrioSeq is compared to CUDA ClustalW \cite{hung2015cuda} and TWILIGHT \cite{tseng2025ultrafast}, two progressive MSA approaches that leverage GPUs for alignment acceleration and quality. For this analysis, we adopt the evaluation methodology from \cite{tseng2025ultrafast}, where experiments on simulated sequence triplets from AliSim \cite{ly2022alisim} with sequence length ranging from 8 to 1024 are performed. For these scenarios, TWILIGHT and TrioSeq were executed on all GPUs, while CUDA ClustalW only runs on NVIDIA A100 and H100 GPUs (due to the lack of support for AMD GPUs). Table \ref{tab:sota} provides the execution times on all tested GPUs in milliseconds for the three solutions, along with TrioSeq's speedups.

Of the three tested approaches for 3-way alignments, CUDA ClustalW has the lowest performance, being at least 25x slower than TWILIGHT in H100 and 61x slower in A100. However, the herein proposed TrioSeq approach outperforms TWILIGHT for triplet alignment, while being an exact algorithm. For the smallest sequence length (8), TrioSeq achieves significant speedups between 15.7x (MI250X) and 83.8x (H100), with speedups decreasing as the sequence length increases, which is to be expected, as TWILIGHT and CUDA ClustalW use pairwise alignments to generate the MSA, which has quadratic complexity, while the exact triplet alignment has cubic complexity. Nevertheless, even with the longest sequence lengths tested, TrioSeq still outperforms TWILIGHT by, at least, 20\% (A100 and H100).

To evaluate the alignment accuracy, FastSP \cite{mirarab2011fastsp} is used to compute errors between inferred and true simulated alignments and calculate the Sum-of-Pairs False Positive (SPFP) and False Negative (SPFN) rates. Alisim \cite{ly2022alisim} is used to generate 10000 simulated alignments with varying lengths, no indel distributions, under the Jukes-Cantor \cite{erickson2010jukes} model with 35\% of invariant sites. For a fair comparison, TrioSeq and TWILIGHT are evaluated on the unaligned sequences using the same scoring model for the alignments. Evaluating all triplet alignments shows that TrioSeq achieves SPFN improvements in 98\% and SPFP improvements in 97\% of the simulated triplets. In particular, in 71\% of the tested triplets, TrioSeq improves SPFN and SPFP between 10\% to 40\%, when compared to TWILIGHT, with improvements up to 80\%.

Finally, to evaluate TrioSeq's performance on real genomic data, an input dataset is generated from \textit{EColi} (accession number SRR35940060\footnote{https://www.ebi.ac.uk/ena/browser/view/SRR35940060}), \textit{Salmonella Enterica} (accession number SRR35984022\footnote{https://www.ebi.ac.uk/ena/browser/view/SRR35984022}), and \textit{Mycobacterium Tuberculosis} (accession number SRR36266169\footnote{https://www.ebi.ac.uk/ena/browser/view/SRR36266169}) Illumina reads. The read length for the first two datasets ranges between 35 and 251, while for the last one, it ranges between 35 and 151. One sequence of each length is sampled from each dataset and triplet combinations are created from the resulting set of reads. In total, the generated dataset has 5509413 sequence triplets to evaluate. When the dataset is evaluated on NVIDIA H100, the fastest tested GPU, TrioSeq achieves a 11x speedup over TWILIGHT when processing only one triplet for iteration. While TWILIGHT processes single sequence triplets, TrioSeq can evaluate all triplets in parallel to reduce the execution time, achieving a speedup of $>$1000x compared to TWILIGHT.

\section{Conclusions}
\label{sec:Conclusions}
Sequence alignment is a key bioinformatics algorithm in various bioinformatics tasks. In particular, MSAs are fundamental for phylogenetic reconstruction, comparative genomics, and epidemiological studies, and their accuracy is essential for downstream analysis. Given the potential of 3-way alignment to increase the quality of MSAs, in this work, TrioSeq was proposed as a novel parallelization approach to achieve unprecedented performance in processing sequence triplets for alignment at the level of single- and multi-GPUs to analyze large-scale datasets. The results on NVIDIA and AMD GPUs showed that TrioSeq achieves at least 20\% better performance when compared to state-of-the-art progressive MSA algorithms, as well as improvements in false negative and false positive rate up to 80\%. Future research includes the implementation of an efficient 3-way alignment for GPU that uses the affine gap model, as well as the alignment traceback to longer sequences, based on Hirschberg's algorithm, and an extension to protein sequences.

\section*{Acknowledgements}
\label{sec:Acknowledgements}
This work was supported by national funds through Fundação para a Ciência e a Tecnologia, I.P. (FCT) under projects UID/50021/2025 (DOI: \url{https://doi.org/10.54499/UID/50021/2025}) and UID/PRR/50021/2025 (DOI: \url{https://doi.org/10.54499/UID/PRR/50021/2025}), LISBOA2030-FEDER-00869000 (2023.18110.ICDT, VERSACOMP, DOI: \url{https://doi.org/10.54499/2023.18110.ICDT}), the UI/BD/154603/2022 research grant, and for awarding us access to Deucalion at MACC, Portugal, through the advanced computing project with reference 2024.14230.CPCA.A3. We also acknowledge the European Union HE Research and Innovation programme under grant agreement No 101092877 (SYCLOPS), and the EuroHPC Joint Undertaking for awarding us access to MareNostrum5 as BSC, Spain, to the EuroHPC supercomputer LUMI, hosted by CSC (Finland), and the LUMI consortium through the advanced computing project with reference EHPC-DEV-2025D04-091.

\bibliographystyle{IEEEtran}
\bibliography{refs}

@article{edgar2022petabase,
  title={Petabase-scale sequence alignment catalyses viral discovery},
  author={Edgar, Robert C and Taylor, Brie and Lin, Victor and Altman, Tomer and Barbera, Pierre and Meleshko, Dmitry and Lohr, Dan and Novakovsky, Gherman and Buchfink, Benjamin and Al-Shayeb, Basem and others},
  journal={Nature},
  volume={602},
  number={7895},
  pages={142--147},
  year={2022},
  publisher={Nature Publishing Group UK London}
}

@article{needleman1970general,
  title={A general method applicable to the search for similarities in the amino acid sequence of two proteins},
  author={Needleman, Saul B and Wunsch, Christian D},
  journal={Journal of molecular biology},
  volume={48},
  number={3},
  pages={443--453},
  year={1970},
  publisher={Elsevier}
}

@article{smith1981identification,
  title={Identification of common molecular subsequences},
  author={Smith, Temple F and Waterman, Michael S and others},
  journal={Journal of molecular biology},
  volume={147},
  number={1},
  pages={195--197},
  year={1981},
  publisher={Elsevier Science}
}

@article{heydari2018browniealigner,
  title={BrownieAligner: accurate alignment of Illumina sequencing data to de Bruijn graphs},
  author={Heydari, Mahdi and Miclotte, Giles and Van de Peer, Yves and Fostier, Jan},
  journal={BMC bioinformatics},
  volume={19},
  number={1},
  pages={311},
  year={2018},
  publisher={Springer}
}

@article{tseng2025ultrafast,
  title={Ultrafast and ultralarge multiple sequence alignments using TWILIGHT},
  author={Tseng, Yu-Hsiang and Walia, Sumit and Turakhia, Yatish},
  journal={Bioinformatics},
  volume={41},
  number={Supplement\_1},
  pages={i332--i341},
  year={2025},
  publisher={Oxford University Press}
}

@article{ye2004approximate,
  title={Approximate multiple protein structure alignment using the sum-of-pairs distance},
  author={Ye, Jieping and Janardan, Ravi},
  journal={Journal of Computational Biology},
  volume={11},
  number={5},
  pages={986--1000},
  year={2004},
  publisher={Mary Ann Liebert, Inc. 2 Madison Avenue Larchmont, NY 10538 USA}
}

@article{hu2021next,
  title={Next-generation sequencing technologies: An overview},
  author={Hu, Taishan and Chitnis, Nilesh and Monos, Dimitri and Dinh, Anh},
  journal={Human immunology},
  volume={82},
  number={11},
  pages={801--811},
  year={2021},
  publisher={Elsevier}
}

@article{daily2016parasail,
  title={Parasail: SIMD C library for global, semi-global, and local pairwise sequence alignments},
  author={Daily, Jeff},
  journal={BMC bioinformatics},
  volume={17},
  number={1},
  pages={81},
  year={2016},
  publisher={Springer}
}

@inproceedings{misra2018performance,
  title={Performance extraction and suitability analysis of multi-and many-core architectures for next generation sequencing secondary analysis},
  author={Misra, Sanchit and Pan, Tony C and Mahadik, Kanak and Powley, George and Vaidya, Priya N and Vasimuddin, Md and Aluru, Srinivas},
  booktitle={Proceedings of the 27th International Conference on Parallel Architectures and Compilation Techniques},
  pages={1--14},
  year={2018}
}

@article{prousalissurvey,
  title={A Survey on Sequence Alignment Algorithms and State-of-the-Art Aligners},
  author={Prousalis, Konstantinos and Georgiou, Konstantinos and Kalogeropoulos, Andreas and Ntalaperas, Dimitrios and Konofaos, Nikos and Aggelis, Lefteris and Papalitsas, Christos and Stavropoulos, Thanos and Gariboldi, Nico},
  journal={ACM Computing Surveys},
  year={2025},
  publisher={ACM New York, NY}
}

@article{maiolo2018progressive,
  title={Progressive multiple sequence alignment with indel evolution},
  author={Maiolo, Massimo and Zhang, Xiaolei and Gil, Manuel and Anisimova, Maria},
  journal={BMC bioinformatics},
  volume={19},
  number={1},
  pages={331},
  year={2018},
  publisher={Springer}
}

@article{rahn2018generic,
  title={Generic accelerated sequence alignment in SeqAn using vectorization and multi-threading},
  author={Rahn, Ren{\'e} and Budach, Stefan and Costanza, Pascal and Ehrhardt, Marcel and Hancox, Jonny and Reinert, Knut},
  journal={Bioinformatics},
  volume={34},
  number={20},
  pages={3437--3445},
  year={2018},
  publisher={Oxford University Press}
}

@article{irawan2015construction,
  title={Construction of phylogenetic tree using neighbor joining algorithms to identify the host and the spreading of SARS epidemic},
  author={Irawan, Mohammad Isa and Amiroch, Siti},
  journal={Journal of Theoretical and Applied Information Technology},
  volume={71},
  number={3},
  pages={424--429},
  year={2015},
  publisher={Little Lion Scientific}
}

@inproceedings{kaur2024construction,
  title={Construction Of Phylogenetic Tree Using UPGMA Method},
  author={Kaur, Amandeep and Kaur, Manjot},
  booktitle={2024 7th International Conference on Contemporary Computing and Informatics (IC3I)},
  volume={7},
  pages={242--246},
  year={2024},
  organization={IEEE}
}

@article{erickson2010jukes,
  title={The jukes-cantor model of molecular evolution},
  author={Erickson, Keith},
  journal={Primus},
  volume={20},
  number={5},
  pages={438--445},
  year={2010},
  publisher={Taylor \& Francis}
}

@inproceedings{muller2022anyseq,
  title={Anyseq/gpu: a novel approach for faster sequence alignment on gpus},
  author={M{\"u}ller, Andr{\'e} and Schmidt, Bertil and Membarth, Richard and Lei{\ss}a, Roland and Hack, Sebastian},
  booktitle={Proceedings of the 36th ACM International Conference on Supercomputing},
  pages={1--11},
  year={2022}
}

@article{mirarab2011fastsp,
  title={FastSP: linear time calculation of alignment accuracy},
  author={Mirarab, Siavash and Warnow, Tandy},
  journal={Bioinformatics},
  volume={27},
  number={23},
  pages={3250--3258},
  year={2011},
  publisher={Oxford University Press}
}

@article{de2016cudalign,
  title={CUDAlign 4.0: Incremental speculative traceback for exact chromosome-wide alignment in GPU clusters},
  author={de Oliveira Sandes, Edans Flavius and Miranda, Guillermo and Martorell, Xavier and Ayguade, Eduard and Teodoro, George and Melo, Alba Cristina Magalhaes},
  journal={IEEE Transactions on Parallel and Distributed Systems},
  volume={27},
  number={10},
  pages={2838--2850},
  year={2016},
  publisher={IEEE}
}

@article{kong2025review,
  title={A Review of Biosequences Alignment, Matching, and Mining Based on GPU},
  author={Kong, Xianghua and Shen, Cong and Tang, Jijun},
  journal={Current Bioinformatics},
  year={2025},
  publisher={Bentham Science Publishers}
}

@article{nobile2017graphics,
  title={Graphics processing units in bioinformatics, computational biology and systems biology},
  author={Nobile, Marco S and Cazzaniga, Paolo and Tangherloni, Andrea and Besozzi, Daniela},
  journal={Briefings in bioinformatics},
  volume={18},
  number={5},
  pages={870--885},
  year={2017},
  publisher={Oxford University Press}
}

@article{ahmed2019gasal2,
  title={GASAL2: a GPU accelerated sequence alignment library for high-throughput NGS data},
  author={Ahmed, Nauman and L{\'e}vy, Jonathan and Ren, Shanshan and Mushtaq, Hamid and Bertels, Koen and Al-Ars, Zaid},
  journal={BMC bioinformatics},
  volume={20},
  number={1},
  pages={520},
  year={2019},
  publisher={Springer}
}

@article{awan2020adept,
  title={Adept: a domain independent sequence alignment strategy for gpu architectures},
  author={Awan, Muaaz G and Deslippe, Jack and Buluc, Aydin and Selvitopi, Oguz and Hofmeyr, Steven and Oliker, Leonid and Yelick, Katherine},
  journal={BMC bioinformatics},
  volume={21},
  number={1},
  pages={406},
  year={2020},
  publisher={Springer}
}

@article{liu2013cudasw++,
  title={CUDASW++ 3.0: accelerating Smith-Waterman protein database search by coupling CPU and GPU SIMD instructions},
  author={Liu, Yongchao and Wirawan, Adrianto and Schmidt, Bertil},
  journal={BMC bioinformatics},
  volume={14},
  number={1},
  pages={117},
  year={2013},
  publisher={Springer}
}

@inproceedings{muller2020anyseq,
  title={AnySeq: A high performance sequence alignment library based on partial evaluation},
  author={M{\"u}ller, Andr{\'e} and Schmidt, Bertil and Hildebrandt, Andreas and Membarth, Richard and Lei{\ss}a, Roland and Kruse, Matthis and Hack, Sebastian},
  booktitle={2020 IEEE International Parallel and Distributed Processing Symposium (IPDPS)},
  pages={1030--1040},
  year={2020},
  organization={IEEE}
}

@article{katoh2010parallelization,
  title={Parallelization of the MAFFT multiple sequence alignment program},
  author={Katoh, Kazutaka and Toh, Hiroyuki},
  journal={Bioinformatics},
  volume={26},
  number={15},
  pages={1899--1900},
  year={2010},
  publisher={Oxford University Press}
}

@article{schmidt2024cudasw++,
  title={CUDASW++ 4.0: ultra-fast GPU-based Smith--Waterman protein sequence database search},
  author={Schmidt, Bertil and Kallenborn, Felix and Chacon, Alejandro and Hundt, Christian},
  journal={BMC bioinformatics},
  volume={25},
  number={1},
  pages={342},
  year={2024},
  publisher={Springer}
}

@article{fei2018fpgasw,
  title={FPGASW: accelerating large-scale Smith--Waterman sequence alignment application with backtracking on FPGA linear systolic array},
  author={Fei, Xia and Dan, Zou and Lina, Lu and Xin, Man and Chunlei, Zhang},
  journal={Interdisciplinary Sciences: Computational Life Sciences},
  volume={10},
  number={1},
  pages={176--188},
  year={2018},
  publisher={Springer}
}

@inproceedings{haghi2021fpga,
  title={An FPGA accelerator of the wavefront algorithm for genomics pairwise alignment},
  author={Haghi, Abbas and Marco-Sola, Santiago and Alvarez, Lluc and Diamantopoulos, Dionysios and Hagleitner, Christoph and Moreto, Miquel},
  booktitle={2021 31st International Conference on Field-Programmable Logic and Applications (FPL)},
  pages={151--159},
  year={2021},
  organization={IEEE}
}

@article{rucci2018swifold,
  title={SWIFOLD: Smith-Waterman implementation on FPGA with OpenCL for long DNA sequences},
  author={Rucci, Enzo and Garcia, Carlos and Botella, Guillermo and De Giusti, Armando and Naiouf, Marcelo and Prieto-Matias, Manuel},
  journal={BMC systems biology},
  volume={12},
  number={Suppl 5},
  pages={96},
  year={2018},
  publisher={Springer}
}

@inproceedings{burchard2023space,
  title={Space efficient sequence alignment for sram-based computing: X-drop on the graphcore IPU},
  author={Burchard, Luk and Zhao, Max Xiaohang and Langguth, Johannes and Bulu{\c{c}}, Ayd{\i}n and Guidi, Giulia},
  booktitle={Proceedings of the International Conference for High Performance Computing, Networking, Storage and Analysis},
  pages={1--16},
  year={2023}
}

@inproceedings{zhao2024ipuma,
  title={iPuma: High-Performance Sequence Alignment on the Graphcore IPU},
  author={Zhao, Max and Burchard, Luk and Schroeder, Daniel Thilo and Langguth, Johannes and Cai, Xing},
  booktitle={ISC High Performance 2024 Research Paper Proceedings (39th International Conference)},
  pages={1--11},
  year={2024},
  organization={Prometeus GmbH}
}

@inproceedings{li2014optimal,
  title={Optimal alignment of three sequences on a gpu},
  author={Li, Junjie and Ranka, Sanjay and Sahni, Sartaj},
  booktitle={Proceedings of the 6th International Conference on Bioinformatics and Computational Biology},
  year={2014}
}

@article{just2001computational,
  title={Computational complexity of multiple sequence alignment with SP-score},
  author={Just, Winfried},
  journal={Journal of computational biology},
  volume={8},
  number={6},
  pages={615--623},
  year={2001},
  publisher={Mary Ann Liebert, Inc.}
}

@article{wallace2006m,
  title={M-Coffee: combining multiple sequence alignment methods with T-Coffee},
  author={Wallace, Iain M and O'sullivan, Orla and Higgins, Desmond G and Notredame, Cedric},
  journal={Nucleic acids research},
  volume={34},
  number={6},
  pages={1692--1699},
  year={2006},
  publisher={Oxford University Press}
}

@inproceedings{sundfeld2013msa,
  title={MSA-GPU: exact multiple sequence alignment using GPU},
  author={Sundfeld, Daniel and de Melo, Alba CMA},
  booktitle={Brazilian Symposium on Bioinformatics},
  pages={47--58},
  year={2013},
  organization={Springer}
}

@article{llados2021accurate,
  title={Accurate consistency-based MSA reducing the memory footprint},
  author={Llad{\'o}s, Jordi and Cores, Fernando and Guirado, Fernando and L{\'e}rida, Josep L},
  journal={Computer Methods and Programs in Biomedicine},
  volume={208},
  pages={106237},
  year={2021},
  publisher={Elsevier}
}

@article{lambert2019impact,
  title={Impact of alignment algorithm on the estimation of pairwise genetic similarity of porcine reproductive and respiratory syndrome virus (PRRSV)},
  author={Lambert, Marie-{\`E}ve and Arsenault, Julie and Delisle, Benjamin and Audet, Pascal and Poljak, Zvonimir and D’Allaire, Sylvie},
  journal={BMC Veterinary Research},
  volume={15},
  number={1},
  pages={135},
  year={2019},
  publisher={Springer}
}

@article{loytynoja2008phylogeny,
  title={Phylogeny-aware gap placement prevents errors in sequence alignment and evolutionary analysis},
  author={Loytynoja, Ari and Goldman, Nick},
  journal={science},
  volume={320},
  number={5883},
  pages={1632--1635},
  year={2008},
  publisher={American Association for the Advancement of Science}
}

@article{seo2022correlations,
  title={Correlations between alignment gaps and nucleotide substitution or amino acid replacement},
  author={Seo, Tae-Kun and Redelings, Benjamin D and Thorne, Jeffrey L},
  journal={Proceedings of the National Academy of Sciences},
  volume={119},
  number={34},
  pages={e2204435119},
  year={2022},
  publisher={National Academy of Sciences}
}

@inproceedings{chien2018three,
  title={Three-dimensional dynamic programming accelerator for multiple sequence alignment},
  author={Chien, Ruei-Ting and Liao, Yi-Lun and Wang, Chien-An and Li, Yu-Cheng and Lu, Yi-Chang},
  booktitle={2018 IEEE Nordic Circuits and Systems Conference (NORCAS): NORCHIP and International Symposium of System-on-Chip (SoC)},
  pages={1--5},
  year={2018},
  organization={IEEE}
}

@inproceedings{chien2022traceback,
  title={Traceback Memory Reduction for Three-Sequence Alignment Algorithm with Affine Gap Models},
  author={Chien, Ruei--Ting and Lin, Mao--Jan and Yeh, Yang--Ming and Lu, Yi--Chang},
  booktitle={2022 Asia-Pacific Signal and Information Processing Association Annual Summit and Conference (APSIPA ASC)},
  pages={1014--1018},
  year={2022},
  organization={IEEE}
}

@article{noah2020major,
  title={Major revisions in arthropod phylogeny through improved supermatrix, with support for two possible waves of land invasion by chelicerates},
  author={Noah, Katherine E and Hao, Jiasheng and Li, Luyan and Sun, Xiaoyan and Foley, Brian and Yang, Qun and Xia, Xuhua},
  journal={Evolutionary Bioinformatics},
  volume={16},
  pages={1176934320903735},
  year={2020},
  publisher={SAGE Publications Sage UK: London, England}
}

@article{capella2013measuring,
  title={Measuring guide-tree dependency of inferred gaps in progressive aligners},
  author={Capella-Guti{\'e}rrez, Salvador and Gabald{\'o}n, Toni},
  journal={Bioinformatics},
  volume={29},
  number={8},
  pages={1011--1017},
  year={2013},
  publisher={Oxford University Press}
}

@article{gotoh1986alignment,
  title={Alignment of three biological sequences with an efficient traceback procedure},
  author={Gotoh, Osamu},
  journal={Journal of Theoretical Biology},
  volume={121},
  number={3},
  pages={327--337},
  year={1986},
  publisher={Elsevier}
}

@article{askari2024three,
  title={Three-Way Alignment Improves Multiple Sequence Alignment of Highly Diverged Sequences},
  author={Askari Rad, Mahbubeh and Kruglikov, Alibek and Xia, Xuhua},
  journal={Algorithms},
  volume={17},
  number={5},
  pages={205},
  year={2024},
  publisher={MDPI}
}

@article{rosenberg2005multiple,
  title={Multiple sequence alignment accuracy and evolutionary distance estimation},
  author={Rosenberg, Michael S},
  journal={Bmc Bioinformatics},
  volume={6},
  number={1},
  pages={278},
  year={2005},
  publisher={Springer}
}

@article{xiao2014giab,
  title={GIAB: Genome reference material development resources for clinical sequencing},
  author={Xiao, Chunlin and Zook, Justin and Trask, Shane and Sherry, Stephen and Genome-in-a-Bottle Consortium},
  journal={Cancer Research},
  volume={74},
  number={19\_Supplement},
  pages={5328--5328},
  year={2014},
  publisher={AACR}
}

@inproceedings{liu2014swaphi,
  title={SWAPHI-LS: Smith-Waterman algorithm on Xeon Phi coprocessors for long DNA sequences},
  author={Liu, Yongchao and Tran, Tuan-Tu and Lauenroth, Felix and Schmidt, Bertil},
  booktitle={2014 IEEE International Conference on Cluster Computing (CLUSTER)},
  pages={257--265},
  year={2014},
  organization={IEEE}
}

@inproceedings{hou2016aalign,
  title={Aalign: A simd framework for pairwise sequence alignment on x86-based multi-and many-core processors},
  author={Hou, Kaixi and Wang, Hao and Feng, Wu-chun},
  booktitle={2016 IEEE International Parallel and Distributed Processing Symposium (IPDPS)},
  pages={780--789},
  year={2016},
  organization={IEEE}
}

@article{carrillo1988multiple,
  title={The multiple sequence alignment problem in biology},
  author={Carrillo, Humberto and Lipman, David},
  journal={SIAM journal on applied mathematics},
  volume={48},
  number={5},
  pages={1073--1082},
  year={1988},
  publisher={SIAM}
}

@inproceedings{de2021multi,
  title={Multi-GPU approach for large-scale multiple sequence alignment},
  author={de O. Siqueira, Rodrigo A and Stefanes, Marco A and Rozante, Luiz CS and Martins-Jr, David C and de Souza, Jorge ES and Araujo, Eloi},
  booktitle={International Conference on Computational Science and Its Applications},
  pages={560--575},
  year={2021},
  organization={Springer}
}

@article{bani2024accelerating,
  title={Accelerating multiple sequence alignments using parallel computing},
  author={Bani Baker, Qanita and Al-Hussien, Ruba A and Al-Ayyoub, Mahmoud},
  journal={Computation},
  volume={12},
  number={2},
  pages={32},
  year={2024},
  publisher={MDPI}
}

@article{carroll2019semiglobal,
  title={Semiglobal sequence alignment with gaps using GPU},
  author={Carroll, Thomas C and Ojiaku, Jude-Thaddeus and Wong, Prudence WH},
  journal={IEEE/ACM Transactions on Computational Biology and Bioinformatics},
  volume={17},
  number={6},
  pages={2086--2097},
  year={2019},
  publisher={IEEE}
}

@inproceedings{ben2020block,
  title={A block-based systolic array on an HBM2 FPGA for DNA sequence alignment},
  author={Ben Abdelhamid, Riadh and Yamaguchi, Yoshiki},
  booktitle={International Symposium on Applied Reconfigurable Computing},
  pages={298--313},
  year={2020},
  organization={Springer}
}

@article{frith2004finding,
  title={Finding functional sequence elements by multiple local alignment},
  author={Frith, Martin C and Hansen, Ulla and Spouge, John L and Weng, Zhiping},
  journal={Nucleic acids research},
  volume={32},
  number={1},
  pages={189--200},
  year={2004},
  publisher={Oxford University Press}
}

@article{nvbio2015,
  title={NVBIO},
  author={Jacopo Pantaleoni and Nuno Subtil},
  year={2015},
  note={\url{https://nvlabs.github.io/nvbio}}
}

@article{notredame2007recent,
  title={Recent evolutions of multiple sequence alignment algorithms},
  author={Notredame, C{\'e}dric},
  journal={PLoS computational biology},
  volume={3},
  number={8},
  pages={e123},
  year={2007},
  publisher={Public Library of Science San Francisco, USA}
}

@article{ly2022alisim,
  title={AliSim: a fast and versatile phylogenetic sequence simulator for the genomic era},
  author={Ly-Trong, Nhan and Naser-Khdour, Suha and Lanfear, Robert and Minh, Bui Quang},
  journal={Molecular biology and evolution},
  volume={39},
  number={5},
  pages={msac092},
  year={2022},
  publisher={Oxford University Press}
}

@article{10002015global,
  title={A global reference for human genetic variation},
  author={1000 Genomes Project Consortium and others},
  journal={Nature},
  volume={526},
  number={7571},
  pages={68},
  year={2015},
  publisher={Nature Publishing Group}
}

@article{colbourn2007lower,
  title={Lower bounds on multiple sequence alignment using exact 3-way alignment},
  author={Colbourn, Charles J and Kumar, Sudhir},
  journal={BMC bioinformatics},
  volume={8},
  number={1},
  pages={140},
  year={2007},
  publisher={Springer}
}

@article{kruspe2007progressive,
  title={Progressive multiple sequence alignments from triplets},
  author={Kruspe, Matthias and Stadler, Peter F},
  journal={BMC bioinformatics},
  volume={8},
  number={1},
  pages={254},
  year={2007},
  publisher={Springer}
}

@article{ogden2006multiple,
  title={Multiple sequence alignment accuracy and phylogenetic inference},
  author={Ogden, T Heath and Rosenberg, Michael S},
  journal={Systematic biology},
  volume={55},
  number={2},
  pages={314--328},
  year={2006},
  publisher={Oxford University Press}
}

@article{yan2013comparative,
  title={A comparative assessment and analysis of 20 representative sequence alignment methods for protein structure prediction},
  author={Yan, Renxiang and Xu, Dong and Yang, Jianyi and Walker, Sara and Zhang, Yang},
  journal={Scientific reports},
  volume={3},
  number={1},
  pages={2619},
  year={2013},
  publisher={Nature Publishing Group UK London}
}

@article{thompson1994clustal,
  title={CLUSTAL W: improving the sensitivity of progressive multiple sequence alignment through sequence weighting, position-specific gap penalties and weight matrix choice},
  author={Thompson, Julie D and Higgins, Desmond G and Gibson, Toby J},
  journal={Nucleic acids research},
  volume={22},
  number={22},
  pages={4673--4680},
  year={1994},
  publisher={Oxford university press}
}

@article{li2003clustalw,
  title={ClustalW-MPI: ClustalW analysis using distributed and parallel computing},
  author={Li, Kuo-Bin},
  journal={Bioinformatics},
  volume={19},
  number={12},
  pages={1585--1586},
  year={2003},
  publisher={Oxford University Press}
}

@article{oliver2005using,
  title={Using reconfigurable hardware to accelerate multiple sequence alignment with ClustalW},
  author={Oliver, Tim and Schmidt, Bertil and Nathan, Darran and Clemens, Ralf and Maskell, Douglas},
  journal={Bioinformatics},
  volume={21},
  number={16},
  pages={3431--3432},
  year={2005},
  publisher={Oxford University Press}
}

@article{hung2015cuda,
  title={CUDA ClustalW: An efficient parallel algorithm for progressive multiple sequence alignment on Multi-GPUs},
  author={Hung, Che-Lun and Lin, Yu-Shiang and Lin, Chun-Yuan and Chung, Yeh-Ching and Chung, Yi-Fang},
  journal={Computational biology and chemistry},
  volume={58},
  pages={62--68},
  year={2015},
  publisher={Elsevier}
}


\end{document}